\documentclass{cimento}
\usepackage{graphicx}
\newcommand{\be}{\begin{equation}}
\newcommand{\ee}{\end{equation}}
\newcommand{\bea}{\begin{eqnarray}}
\newcommand{\eea}{\end{eqnarray}}
\newcommand{\bean}{\begin{eqnarray*}}
\newcommand{\eean}{\end{eqnarray*}}
\title{The role of the screen factor in GRBs}
\author{Giuliano Preparata\from{ins:mi}Remo Ruffini\from{ins:icra}She-Sheng Xue\from{ins:icra}
} 
\instlist{\inst{ins:mi}
Department of Physics, University of Milan, Via Celoria 16, 20133, Milan - Italy\\
\inst{ins:icra} ICRA - International Center for Relativistic Astrophysics, Dipartimento di Fisica, University of Rome "la Sapienza", P.le A. Moro, 5, 00185, Rome - Italy \\ ruffini@icra.it}
\begin{document}
\maketitle
\begin{abstract}
We derive the screen factor for the radiation flux from an optically thick plasma of electron-positron pairs and photons, created by vacuum polarization process around a black hole endowed with electromagnetic structure.
\end{abstract}

\section{On the screen factor}

In our model for Gamma Ray Bursts, based on the electromagnetic black hole (EMBH) creating a dyadosphere\cite{prx}, there is a plasma made by electron-positron pairs and photons with extremely high density ($\sim 10^{30}/{\rm cm^3}$). This plasma relativistically expands outward from EMBH as soon as it is formed. This relativistically expanding plasma of electron-positron pairs and photons evolves with a slab-like profile of constant thickness in the laboratory frame: a pair electromagnetic pulse (PEM - pulse)\cite{rswx}. In the comoving frame of the PEM pulse, the plasma of electron-positron pairs and photons is assumed to be in thermal equilibrium and characterized by the energy density ($\rho$), the number density $(n_{e^+e^-})$ and the temperature $T$. Since the pair-number density is very high, the plasma is optically thick reaching the transparency condition only at the surface. In this paper, we evaluate the photon flux $F$ radiated from such a plasma.

The photon flux radiated results from the photon transport inside the plasma due to diffusion phenomenon caused by a gradient temperature in the radial direction, the diffusion constant is assumed to be independent of photon frequency and the photon flux in such radiative equilibrium is independent of time.
In order to have a non-vanishing photon flux in the radiative equilibrium, there must be a non-vanishing temperature gradient. For this, we assume that the temperature distribution in the radial direction is
\begin{equation}
T=T_c(1-\frac{r^2}{R^2}).
\label{td}
\end{equation}

We separate the plasma into two regions, the core and the halo: (i) the core is optically thick and has average density $\bar\rho$ and (ii) the halo is transparent and has a temperature
\begin{equation}
T_h=T_c\left(1-\left(\frac{R-\Delta}{R}\right)^2\right)\simeq T_c\left(\frac{2\Delta}{R}\right),
\label{ht}
\end{equation}
where $\Delta$ is the thickness of the halo and $T_h$ is the surface temperature of the plasma.The total flux $F$ at the surface of the plasma is, 
\begin{equation}
F\sim 4\pi R^2 \sigma T_h^4,
\label{sl}
\end{equation}
where $\sigma$ is the Stefan-Boltzmann constant. From Eq.(\ref{ht}) in Eq.(\ref{sl}),  we obtain the flux (\ref{sl}) in terms of halo thickness:
\begin{equation}
F\sim  4\pi R^2\sigma T_h^4\simeq R^2 \sigma T^4_c\left({2\Delta\over R}\right)^4.
\label{ht2}
\end{equation}

We can evaluate as well the flux from the radiation transfer equation (see e.g. \cite{book1}):
\begin{equation}
F\sim 4\pi\sigma\frac{r^3}{R^2}T_cT(r)^3L_\gamma\sim 4\pi R \sigma T_c^4 L_\gamma.
\label{slt}
\end{equation}
where $L_\gamma = \frac{1}{\sigma_T\bar n_{e^+e^-}}$ is the mean free path of the photons inside the plasma, $\sigma_T$ is the Thomson cross-section and $T(r)$ is the temperature at depth $r$. 

From these two estimates of the total flux, (\ref{ht2}) and (\ref{slt}), we have:
\begin{eqnarray}
R^2T_c^4\left({2\Delta\over R}\right)^4&\sim& T_c^4RL_\gamma,\nonumber\\
\left({2\Delta\over R}\right)^4&\sim& \left({L_\gamma\over R}\right)
\label{slt3}
\end{eqnarray}
Thus, we obtain
\begin{equation}
{2\Delta\over R}\sim \left({L_\gamma\over R}\right)^{1\over4},\hskip0.5cm \Delta ={R\over2}\left({L_\gamma\over R}\right)^{1\over 4}.
\label{end}
\end{equation}
Substituting Eq.(\ref{end}) in Eq.(\ref{ht2}) we obtain:
\begin{equation}
F\sim  4\pi R^2\sigma T^4_c\frac{L_\gamma}{R}
\label{eqf}
\end{equation}
Now, $F\sim  4\pi R^2 \sigma T^4_c$ is the flux emitted by the core of the plasma if optically thin, so the suppression factor is
\begin{equation}
S\simeq \frac{L_\gamma}{R}.
\label{s99}
\end{equation}

We can as well obtain the analogous results from the diffusion equation for the photon energy-density $\rho$ as:
\begin{equation}
{\partial\rho\over\partial \tau} +\lambda
(\rho-\rho_\circ)
=D\nabla^2\rho,
\label{epr'}
\end{equation}
where the diffusion constant $D={cL_\gamma\over 3}$ and the relaxation time of photon diffusion is ${1\over\lambda}\simeq {R\over v}$, where $v$, the velocity of photon diffusion, can be calculated by a random walk process (see e.g. \cite{book1}):
\begin{equation}
v\simeq {2\over3}{L_\gamma\over R}c.
\label{pd}
\end{equation}
In Eq.(\ref{epr'}), $\rho(r,t)$ is the value of the photon energy density and $\rho_\circ(r)$ is the photon energy density at radiative equilibrium. Such energy density distribution is assumed of the form
\begin{equation}
\rho_\circ(r)=\rho_\circ(1-\frac{r^n}{R^n})\simeq \rho_\circ\theta (R-r).
\label{tdt}
\end{equation}

Since the results is very weekly dependent on the index n. With the damped heat-kernel we have the solution:
\begin{equation}
\rho={1\over\lambda}\int^t_0dt'e^{-\lambda(t-t')}{e^{-{
(\vec x -\vec
x')^2\over4D(t-t')}}\over[4\pi D(t-t')]^{3\over2}}
\rho_\circ\theta 
(R-r') d^3x',
\label{pp}
\end{equation}
which can be written as 
\begin{equation}
\rho= -{\lambda\over\sqrt{\pi}}\rho_\circ\int^t_0d\tau 
e^{\lambda\tau}
\int^{r-R\over\sqrt{4D\tau}}_{r\over\sqrt{4D\tau}} e^{-\xi^2}d\xi.
\label{ppp}
\end{equation}
We compute the photon current for $r\simeq R$ and $t\rightarrow\infty$, 
\begin{eqnarray}
j_r&=&-D\left({\partial\rho\over\partial r}\right)_{r=R}
={\lambda\over2\sqrt{\pi}}\rho_\circ D\int^t_0d\tau 
e^{-\lambda\tau}{1\over(2D\tau)^{1\over2}},\nonumber\\
&=&{\rho_\circ\lambda\over2\sqrt{\pi}} \left({D\over\lambda}\right)^{1\over2}\int^\infty_0dx 
e^{-x}{1\over x^{1\over2}}
\label{pppp}
\end{eqnarray}
and
\begin{equation}
j_r\rightarrow \rho_\circ {\Gamma(-{1\over2})\over2\sqrt{\pi}}\left(
\lambda D\right)^{1\over2}={\rho_\circ c\over2}\left[{v L_\gamma
\over3cR}\right]^{1\over2}.
\label{sf}
\end{equation}
Comparing with the photon current computed in an optical thin medium $j_r={\rho_\circ c\over2}$, we obtain the screening 
factor
\begin{equation}
S={\sqrt{2}\over3}\frac{L_\gamma}{R}.
\label{s1}
\end{equation}
in substantial agreement with Eq.(\ref{s99})   

\section{\bf  The observable radiation flux}\label{flux}

We can now estimate the observed radiation flux from the PEM pulse assuming that the  photons are in equilibrium at the same temperature $T$ with electron-positron pairs before they decouple. In the comoving frame of photons and electron-positron pairs plasma fluid, the black-body spectrum of photons that are in thermal with $e^+e^-$-pairs is given by,
\begin{equation}
{dn_\gamma\over d^3k_c}= {1\over \pi^2\hbar^3}{1\over \exp\left({\omega_\gamma(|{\bf k}_c|)\over kT}\right)-1},
\label{cspectrum}
\end{equation}
where $n_\gamma$ is the number-density of photons, $T$ is the temperature and $\omega_\gamma,{\bf k}_c$ are energy-momentum in the comoving frame. Integrating over all photon momenta, we obtain the black body radiation flux in the comoving frame(egrs/sec./4$\pi$),
\begin{equation}
F_c=aT^4\tilde R^2c,
\label{fc}
\end{equation}
where $a$ is the Steven-Boltzmann constant and $\tilde R$ is the radius of the front of the plasma fluid.

The energy and momentum of photons in its comoving frame and local laboratory frame are related,
\begin{eqnarray}
\omega_\gamma &=&\bar\gamma E_\gamma(1-{v\over c}\cos\theta),\hskip0.3cm \omega_\gamma =|{\bf k}_c|,
\label{entran}\\
{\bf k}_c&=&-|{\bf k}|\sqrt{1-\cos^2\theta}{\bf u}+\bar\gamma |{\bf k}|(\cos\theta-{v\over c}){\bf v},\label{kntran}\\
{\bf k}&=&E_\gamma(-\sin\theta{\bf u} +\cos\theta{\bf v})\hskip0.3cm  |{\bf k}|=E_\gamma,
\label{kdirection}
\end{eqnarray}
where $E, {\bf k}$ are energy-momentum in the local laboratory frame, $\theta$ is the angle (in the laboratory frame) between the radial expanding-velocity and direction from the origin of the plasma fluid to the observer, ${\bf v}$ is a unit vector along the radial expanding-velocity of the plasma fluid and ${\bf u}$ is a unit vector
transverse to ${\bf v}$. In the comoving frame, shown by Eq.(\ref{kntran}), photons radiating out of the plasma fluid must have
\begin{equation}
\cos\theta\ge {v\over c},
\label{cos}
\end{equation}
so that the component of the photon momentum in the radial expanding-velocity direction is positive.

Note that due to the Louiville theorem (the phase invariance) (see e.g. \cite{eh}, the spectrum of photons in the laboratory frame has the same as Eq.(\ref{cspectrum}) by replacing $\omega_\gamma\rightarrow E_\gamma$, $T\rightarrow T_{\rm lab}$. Considering Eq.(\ref{entran}), we have
\begin{equation}
{dn_\gamma\over d^3k}= {1\over \pi^2\hbar^3}{1\over \exp\left({E_\gamma(|{\bf k}|)\bar\gamma (1-{v\over c}\cos\theta)\over kT}\right)-1},
\label{lspectrum}
\end{equation}
which leads to the temperature $T_{\rm lab}$ in the local laboratory frame,
\begin{equation}
T_{\rm lab}={T\over\bar\gamma (1-{v\over c}\cos\theta)}.
\label{tspectrum}
\end{equation}
Using eq.(\ref{fc}), we obtain the black body radiation flux in the laboratory frame,
\begin{equation}
F_{BB}={1\over \left[\gamma (1-{v\over c}\cos\theta)\right]^4}F_c.
\label{f}
\end{equation}

Integrating over the $\cos\theta$ in the range of $(\ref{cos})$, for $\gamma\gg 1$ we obtain the total black-body radiation flux per unit solid angle is approximately given by (ergs/second) 
\begin{equation}
F_{BB}\simeq 4\pi\sigma T^4\tilde R^2c\gamma^2,
\label{bssc1}
\end{equation}
where $\gamma$ is the Lorentz factor. The screened radiation flux from the PEM pulse is then given by (ergs/second)
\begin{equation}
\tilde F_{BB}\simeq 4\pi\sigma T^4\tilde R^2c\gamma^2 S.
\label{bssc}
\end{equation}
with $S$ given by Eq.(\ref{s1}).

\end{document}